\documentclass[aps,pre,
twocolumn
]{revtex4-1}

\usepackage{mathrsfs}
\usepackage{bm}
\usepackage{amsmath}
\usepackage{amssymb}
\usepackage{graphicx}
\usepackage{mathrsfs}
\usepackage{MnSymbol}

\makeatletter

\usepackage{hyperref}
\usepackage{color}
\definecolor{myblue}{RGB}{0,50,200}
\hypersetup{
colorlinks,
citecolor=myblue,
linkcolor=myblue,
urlcolor=myblue
}

\allowdisplaybreaks

\newcommand{\mca}{\mathcal}
\newcommand{\mbb}{\mathbb}

\newcommand{\avg}[1]{\langle #1\rangle}

\newcommand{\set}[1]{\lbrace #1\rbrace}
\newcommand{\bra}[1]{\left( #1 \right)}
\newcommand{\bras}[1]{\left[ #1 \right]}

\newcommand{\pp}{\partial}
\newcommand{\Var}[1]{\avg{\avg{#1}}}
\newcommand{\mX}{\mathcal{X}}
\newcommand{\mY}{\mathcal{Y}}

\begin{document}
\title{Thermodynamic uncertainty relations under arbitrary control protocols}

\author{Tan Van Vu}
\email{tan@biom.t.u-tokyo.ac.jp}

\affiliation{Department of Information and Communication Engineering, Graduate
School of Information Science and Technology, The University of Tokyo,
Tokyo 113-8656, Japan}

\author{Yoshihiko Hasegawa}
\email{hasegawa@biom.t.u-tokyo.ac.jp}

\affiliation{Department of Information and Communication Engineering, Graduate
School of Information Science and Technology, The University of Tokyo,
Tokyo 113-8656, Japan}

\date{\today}

\begin{abstract}
Thermodynamic uncertainty relations quantifying a trade-off between current fluctuation and entropy production have been found in various stochastic systems.
Herein, we study the thermodynamic uncertainty relations for Langevin systems driven by an external control protocol.
Using information-theoretic techniques, we derive the uncertainty relations for arbitrary observables satisfying a scaling condition in both overdamped and underdamped regimes.
We prove that the observable fluctuation is constrained by both entropy production and a kinetic term.
The derived bounds are applicable to both current and noncurrent observables, and hold for arbitrary time-dependent protocols, thus, providing a wide range of applicability.
We illustrate our universal bounds with the help of three systems: a dragged Brownian particle, a Brownian gyrator, and a stochastic underdamped heat engine.
\end{abstract}

\maketitle

\section{Introduction}
Stochastic thermodynamics \cite{Jarzynski.1997.PRL,Sekimoto.1998.PTPS,Seifert.2012.RPP} provides a rigorous framework for investigating the physical properties of small systems.
On theoretical grounds, it is known that thermodynamic costs place fundamental limits on the performance of real-world systems, ranging from living organisms to artificial devices.
Investigating such trade-off relations provides insights into the optimal design principles for such systems.

In recent years, powerful inequalities known as thermodynamic uncertainty relations (TURs) have been discovered for nonequilibrium systems \cite{Barato.2015.PRL,Gingrich.2016.PRL}.
They assert a trade-off between the current fluctuation and dissipation quantified via entropy production; i.e., a high precision of currents is unattainable without increasing the associated entropy production.
Originally, TURs imposed the following bound in steady-state systems:
\begin{equation}
\frac{\avg{\phi}^2}{\Var{\phi}}\le\frac{\avg{\sigma}}{2},\label{eq:con.TUR}
\end{equation}
where $\phi$ is an arbitrary time-integrated current, $\avg{\phi}$, $\Var{\phi}:=\avg{\phi^2}-\avg{\phi}^2$ are its mean and variance, respectively, and $\avg{\sigma}$ is the average entropy production.
This bound was first derived for biomolecular processes \cite{Barato.2015.PRL} and later proven for continuous-time Markov jump processes \cite{Gingrich.2016.PRL,Horowitz.2017.PRE} and overdamped Langevin systems \cite{Pigolotti.2017.PRL,Dechant.2018.JSM}.
Subsequently, the violation of the original bound has been found for other dynamics, e.g., for discrete-time Markov chains \cite{Proesmans.2017.EPL}, transport systems \cite{Agarwalla.2018.PRB}, and underdamped dynamics \cite{Chun.2019.PRE,Vu.2019.PRE.Underdamp}.
TURs have been intensively refined in other contexts that include both classical and quantum systems \cite{Pietzonka.2016.PRE,Polettini.2016.PRE,Garrahan.2017.PRE,Barato.2018.NJP,Macieszczak.2018.PRL,Brandner.2018.PRL,Hasegawa.2019.PRE,Koyuk.2019.PRL,Terlizzi.2019.JPA,Vu.2019.PRE.Underdamp,Barato.2019.JSM,Gupta.2019.arxiv}.
A remarkable application of TURs is the estimation of entropy production \cite{Li.2019.NC}.
By observing various fluctuating currents, a lower bound on entropy production can be inferred.

In this paper, we focus on extending the applicability of TURs that have been derived for currents in steady-state systems.
Considering general Langevin systems driven by an arbitrary time-dependent control protocol, we derive uncertainty relations for both current and noncurrent observables that satisfy a scaling condition.
We prove for both overdamped and underdamped systems that the observable fluctuation is bounded by entropy production and a kinetic term.
Notably, the derived bounds are not static; however, they are dynamic with respect to the observables and are tighter than the original for a broad class of observables.
When the target system is unidirectionally affected by other systems, we prove a tighter bound for observables in the target system.
That is, fluctuations of the observables are not constrained by the dissipation costs of other systems but by the information flows between them and the target system.
Our results allow the investigation of arbitrary Langevin systems, ranging from relaxation processes to externally controlled systems such as stochastic heat engines.
We apply the results to investigate three systems: a dragged Brownian particle, a Brownian gyrator, and a stochastic underdamped heat engine.

Recent studies have made advances in generalizing TURs.
It has been shown that a TUR is a direct consequence of the detailed fluctuation theorem, regardless of the underlying dynamics \cite{Hasegawa.2019.PRL,Vu.2019.arxiv.Feedback,Timpanaro.2019.PRL}.
This bound is more applicable than the original, i.e., it holds for arbitrary currents and arbitrary dynamics as long as the fluctuation theorem is provided.
However, it pays the cost of a weaker predictive power.
A generalization to systems with a broken symmetry, known as the hysteretic TUR, has been conducted \cite{Proesmans.2019.JSM,Potts.2019.PRE}.
This bound requires the evaluation of currents and entropy production in the backward process, having the following form:
\begin{equation}
\frac{(\avg{\phi}+\avg{\phi}_{\rm b})^2}{\Var{\phi}+\Var{\phi}_{\rm b}}\le\exp\bra{\frac{\avg{\sigma}+\avg{\sigma}_{\rm b}}{2}}-1,\label{eq:hys.TUR}
\end{equation}
where $\avg{\cdot\cdot}_{\rm b}$ denotes averages taken over an ensemble in the backward experiment.
Another extension that holds for arbitrary dynamics reads \cite{Vu.2019.PRE.Delay,Falasco.2019.arxiv}
\begin{equation}
\frac{\avg{\phi}^2}{\Var{\phi}}\le\frac{e^{\avg{\tilde{\sigma}}}-1}{2},\label{eq:rev.TUR}
\end{equation}
where $\avg{\tilde{\sigma}}$ is the Kullback--Leibler divergence between distributions of the forward path and its reversed counterpart in the system.
However, $\avg{\tilde{\sigma}}$ is not equal to the entropy production $\avg{\sigma}$, except in steady-state systems with time-reversal symmetry.
Despite the generalities of Eqs.~\eqref{eq:hys.TUR} and \eqref{eq:rev.TUR}, it is difficult to infer the entropy production from these bounds.

\section{Main results}
For the sake of simplicity, we will describe our results with one-dimensional systems.
The generalization to multidimensional systems is straight-forward.
Unlike in most previous studies, where the system is assumed to be in a stationary or in a transient regime, here, the system starts from an arbitrary distribution at time $t=0$ and is subsequently driven by an external control protocol $\lambda$ up to time $t=\tau$.
When $\lambda$ is absent, it becomes a relaxation process.
Let $\Gamma$ denote the trajectory of system states during this time interval and $\phi(\Gamma)$ be a trajectory-dependent observable which can be time-symmetric.
We aim to derive a bound on the relative fluctuation of $\phi(\Gamma)$.

We consider observables that satisfy the scaling condition: $\phi(\theta\Gamma)=\theta^\kappa\phi(\Gamma)$ for some constant $\kappa>0$ and for all $\theta\in\mbb{R}$.
Given a trajectory $\Gamma=[x(t)]_{t=0}^{t=\tau}$, this can be satisfied with a current $\phi(\Gamma)=\int_{0}^{\tau}dt\,x^{\kappa-1}\circ\dot{x}$ or a noncurrent observable $\phi(\Gamma)=\int_{0}^{\tau}dt\,x^\kappa$.
Here, $\circ$ denotes the Stratonovich product and the dot indicates the time derivative.
Moreover, $\phi$ can be a discrete-time observable, e.g., $\phi(\Gamma)=\sum_{i}c_ix(t_i)^\kappa$, where $0\le t_i\le\tau$ is the predetermined time and $c_i$ is an arbitrary coefficient.
From a practical perspective, measurements are discretely performed in most cases; thus, the acquisition of continuous-time observables may be difficult.
Consequently, a bound on such discrete-time observables provides a useful tool with regard to thermodynamic inference problems.
It is noteworthy that these noncurrent observables cannot be applied with the TURs previously reported.
Hereafter, we consider these three types of observables.

\subsection{Bounds for a full system}
First, we consider a general overdamped Langevin system, whose dynamics are governed by the following equation:
\begin{equation}
\dot{x}=F(x,\lambda)+\xi,
\end{equation}
where $F(x,\lambda)$ is the total force and $\xi$ is a zero-mean Gaussian white noise with variance $\avg{\xi(t)\xi(t')}=2D\delta(t-t')$.
Here, $D>0$ denotes the noise intensity.
Throughout this work, Boltzmann's constant is set to $k_{\rm B}=1$.
Let $\rho(x,t)$ denote the probability distribution function of the system being in state $x$ at time $t$.
Then, its time evolution can be described using the Fokker-Planck equation as $\pp_t\rho(x,t)=-\pp_xj(x,t)$, where $j(x,t)=F(x,\lambda)\rho(x,t)-D\pp_x\rho(x,t)$ is the probability current.
The dynamical solution of this differential equation is uniquely determined if the initial distribution $\rho(x,0)=\rho_{\rm i}(x)$ is given.

As our first main result, we prove that the observable fluctuation is bounded as
\begin{equation}
\frac{\avg{\phi}^2}{\Var{\phi}}\le \frac{1}{\kappa^2}\bra{2\avg{\sigma}+\chi_{\rm o}+\psi_{\rm o}},\label{eq:odp.TUR}
\end{equation}
where $\chi_{\rm o}:=\avg{\int_0^\tau dt\,\Lambda_{\rm o}(x,t)}$ is a kinetic term and $\psi_{\rm o}:=\avg{(x\pp_x\rho_{\rm i}/\rho_{\rm i})^2}_{\rho_{\rm i}}-1$ is a nonnegative boundary value that can be neglected for long observation times.
Here, $\Lambda_{\rm o}=[(\pp_x[xF])^2-4F\pp_x(xF)-4D\pp_x^2(xF)]/2D$ is a function in terms of force and position.

Next, we consider a general underdamped Langevin system, where inertial effects cannot be neglected.
The system consists of a particle being in contact with an equilibrium heat bath.
Its dynamics are described by the following equations:
\begin{equation}
\dot{x}=v,~m\dot{v}=-\gamma v+F(x,\lambda)+\xi,
\end{equation}
where $m$, $\gamma$ are the mass and friction coefficient of the particle, respectively.
Let $\rho(x,v,t)$ be the phase-space probability distribution function of the system at time $t$.
Assuming that the system evolves from an initial distribution $\rho(x,v,0)=\rho_{\rm i}(x,v)$;
then, $\rho(x,v,t)$ follows the Fokker-Planck equation, $\pp_t\rho(x,v,t)=-\pp_xj_x(x,v,t)-\pp_vj_v(x,v,t)$, where $j_x(x,v,t)=v\rho(x,v,t)$ and $j_v(x,v,t)=1/m[-\gamma v+F(x,\lambda)-D/m\pp_v]\rho(x,v,t)$ are probability currents.
Since the position $x$ and the velocity $v$ are the freedom degrees of the system, the trajectory can be written as $\Gamma=[x(t),v(t)]_{t=0}^{t=\tau}$.

For observables that satisfy the scaling condition, we prove that
\begin{equation}
\frac{\avg{\phi}^2}{\Var{\phi}}\le \frac{1}{\kappa^2}\bra{2\avg{\sigma}+\chi_{\rm u}+\psi_{\rm u}},\label{eq:udp.TUR}
\end{equation}
where $\chi_{\rm u}:=\avg{\int_0^\tau dt\,\Lambda_{\rm u}(x,v,t)}$ is a kinetic term and $\psi_{\rm u}:=\avg{[(x\pp_x\rho_{\rm i}+v\pp_v\rho_{\rm i})/\rho_{\rm i}]^2}_{\rho_{\rm i}}-4$ is a nonnegative boundary term that can be neglected for large $\tau$.
Here, $\Lambda_{\rm u}=[\bra{F-x\pp_xF}^2-4\gamma^2v^2+8\gamma D/m]/2D$.
Inequality \eqref{eq:udp.TUR} is our second main result.
The detailed derivations of the bounds and the generalization to the multidimensional case are presented in Appendix \ref{app:deriv.fullsys} and \ref{app:multidimen}, respectively.

We make several remarks about our main results [Eqs.~\eqref{eq:odp.TUR} and \eqref{eq:udp.TUR}].
These inequalities hold for arbitrary protocol $\lambda$, arbitrary initial distribution $\rho_{\rm i}$, and finite observation time $\tau$; thus, they are also valid for steady-state systems.
Interestingly, the derived bounds involve the scaling power $\kappa$; as $\kappa$ is large enough, the bounds become tighter than the original [Eq.~\eqref{eq:con.TUR}].
Moreover, unlike the reported bounds, which deal only with currents, our bounds are applicable for current, noncurrent, and discrete-time observables, as well as for linear combinations of these observables with the same scaling power.

In addition to entropy production, the bounds contain kinetic terms $\chi_{\set{\rm o,\,u}}$.
They are averages of observables, which can be calculated based on the observed trajectories.
As will be shown later, these terms play an important role in the bounds, i.e., the observable fluctuation cannot be solely bounded by the entropy production, even with the exponential bound $(e^{\avg{\sigma}}-1)/2$.
Furthermore, the fluctuation of a noncurrent observable, $\avg{\phi}^2/\Var{\phi}$, may not vanish in equilibrium, for example, for $\phi(\Gamma)=\int_{0}^{\tau}dt\,x^2$, while entropy production always does, i.e., $\avg{\sigma}=0$.
In this scenario, $\chi_{\set{\rm o,\,u}}$ are the key quantities that constrain fluctuations of such noncurrent observables.

We provide an intuitive explanation regarding why kinetic terms appear in the bounds.
Entropy production, which is quantified via irreversible currents of probability density, characterizes the strength of the currents in the system.
Zero entropy production implies that there is no current in the system.
Therefore, its genuine contribution to the bounds is the constraint on fluctuations of currents.
To constrain fluctuations of noncurrent components (e.g., time-symmetric changes), another complement to entropy production, which is identified here as $\chi$, is necessary \cite{Maes.2017.PRL}.

With respect to attaining the derived bounds, one can show that the exact equality condition cannot be attained (from the equality condition of the Cauchy--Swartz inequality) \cite{Vu.2019.PRE.Underdamp}.
Unlike original TUR, which is saturated near equilibrium \cite{Hasegawa.2019.PRE}, attaining the derived bounds is not ensured in such regimes.
As shown later, the bounds become tight for long observation times.
For short observation times, a gap exists between the fluctuations of the observables and the derived bounds.
This occurs due to the dominance of the boundary terms, $\psi_{\rm o,u}$.

\subsection{Bounds for a subsystem}
\begin{figure}[t]
	\centering
	\includegraphics[width=8.5cm]{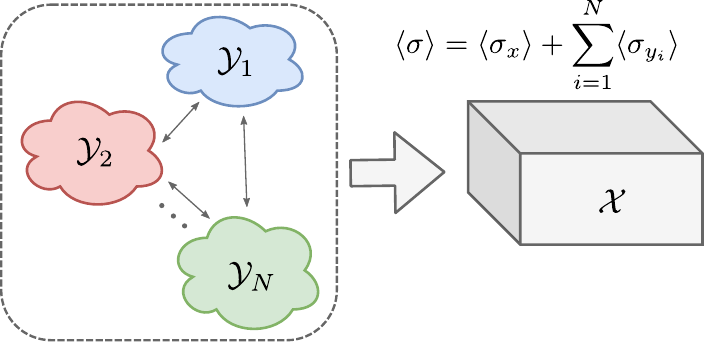}
	\protect\caption{Illustration of multipartite interacting systems described by Langevin equations. The target system $\mX$ is unidirectionally affected by other systems $\mY_1,\dots,\mY_N$. The total entropy production $\avg{\sigma}$ of the full system (including $\mX,\mY_1,\dots,\mY_N$) can be decomposed into nonnegative components as $\avg{\sigma}=\avg{\sigma_x}+\sum_{i=1}^N\avg{\sigma_{y_i}}$, in which each component represents the contribution of the entropy production from each corresponding system.}\label{fig:interacting.systems}
\end{figure}

Let us consider a situation where the target system $\mX$ undergoes unidirectional interactions with other systems $\mY_1,\dots,\mY_N$.
Each of these systems is coupled to a distinct thermal reservoir.
We assume that $\mX$ does not affect the dynamics of $\mY_i$ for all $i$, while each system $\mY_i$ can arbitrarily affect other systems (see Fig.~\ref{fig:interacting.systems} for the illustration).
An example of this situation is a system that involves continuous-time imperfect measurement and feedback control.
The systems $\{\mY_i\}$ correspond to the errors in the readout when performing measurements on $\mX$.
These errors instantly affect the system dynamics of $\mX$ via feedback control.
For simplicity, we consider the case $N=1$, and the system states of $\mX$ and $\mY$ are represented by variables $x$ and $y$, respectively.
The dynamics of $\mX$ and $\mY$ are governed by the following equations:
\begin{align}
\dot{x}&=F_x(x,y,\lambda)+\xi_x,\\
\dot{y}&=F_y(y,\lambda)+\xi_y,
\end{align}
where $\xi_x$ and $\xi_y$ are uncorrelated zero-mean Gaussian white noises satisfying $\avg{\xi_z(t)\xi_z(t')}=2D_z\delta(t-t')$, for each $z\in\{x,y\}$.
Let $\rho_{\rm i}(x,y):=\rho(x,y,0)$ denote the distribution function at time $t=0$.
The corresponding Fokker--Planck equation reads $\pp_t\rho(x,y,t)=-\sum_{z\in\{x,y\}}\pp_zj_z(x,y,t)$, where $j_z=F_z\rho-D_z\pp_z\rho$ is the probability current.

The entropy production of the full system (including both $\mX$ and $\mY$) is given by
\begin{equation}
\avg{\sigma}=\int_0^\tau dt\int dxdy\bra{\frac{j_x(x,y,t)^2}{D_x\rho(x,y,t)}+\frac{j_y(x,y,t)^2}{D_y\rho(x,y,t)}}.
\end{equation}
It can be rewritten as $\avg{\sigma}=\avg{\sigma_x}+\avg{\sigma_y}$, where $\avg{\sigma_z}=\int_0^\tau dt\int dxdy\,j_z(x,y,t)^2/D_z\rho(x,y,t)\ge 0$ is the entropy production contribution from $z\in\{x,y\}$.
Moreover, $\avg{\sigma_x}$ can be further decomposed as $\avg{\sigma_x}=\avg{\Delta s_x}+\avg{q_x}/D_x+I_{x}$ \cite{Horowitz.2015.JSM}, where $\avg{\Delta s_x}$ denotes the change in the Shannon entropy of $\mX$, $\avg{q_x}$ denotes the heat flow from $\mX$ to the reservoir, and $I_{x}$ represents information flow into and out of $\mX$.
Note that because of information exchanged between $\mX$ and $\mY$, $\avg{\Delta s_x}+\avg{q_x}/D_x$ can be negative.

For arbitrary observables involving only $x$ and satisfying the scaling condition with the scaling power $\kappa$, we can prove that
\begin{equation}
\frac{\avg{\phi}^2}{\Var{\phi}}\le\frac{1}{\kappa^2}(2\avg{\sigma_x}+\chi_x+\psi_x),\label{eq:inter.TUR}
\end{equation}
where $\chi_x:=\avg{\int_0^\tau dt\,\Lambda_x(x,y,t)}$ is a kinetic term and $\psi_x:=\avg{(x\pp_x\rho_{\rm i}/\rho_{\rm i})^2}_{\rho_{\rm i}}-1$ is a nonnegative boundary value.
Here, $\Lambda_x=[(\pp_x[xF_x])^2-4F_x\pp_x(xF_x)-4D_x\pp_x^2(xF_x)]/2D_x$.
Inequality \eqref{eq:inter.TUR} is our third main result.
Note that the bound in Eq.~\eqref{eq:odp.TUR} also holds for this situation when we consider a full system that includes both $\mX$ and $\mY$.
The difference between these bounds [Eqs.~\eqref{eq:odp.TUR} and \eqref{eq:inter.TUR}] is that the former contains terms that originate from all the systems, i.e., $\mX$ and $\mY$, while the latter only involves contributions from $\mX$.
Focusing on the entropy production term, $\avg{\sigma_x}$, the bound in Eq.~\eqref{eq:inter.TUR} implies that the fluctuations of the observables in $\mX$ are not constrained by the dissipation in $\mY$ but by that in $\mX$, $\avg{\Delta s_x}+\avg{q_x}/D_x$, and the information flow between $\mX$ and $\mY$, $I_x$.
This implication agrees with intuition.
In general, the bound in Eq.~\eqref{eq:inter.TUR} is tighter than that in Eq.~\eqref{eq:odp.TUR} and reduces to the same one when there is no interaction between $\mX$ and $\mY$.
A bound for underdamped systems can be analogously derived.
The derivation of Eq.~\eqref{eq:inter.TUR} is provided in Appendix \ref{app:deriv.subsys}.

\section{Examples}
In this section, we illustrate our results with the help of three systems, as follows.

\subsection{Dragged Brownian particle}
\begin{figure}[t]
	\centering
	\includegraphics[width=8.5cm]{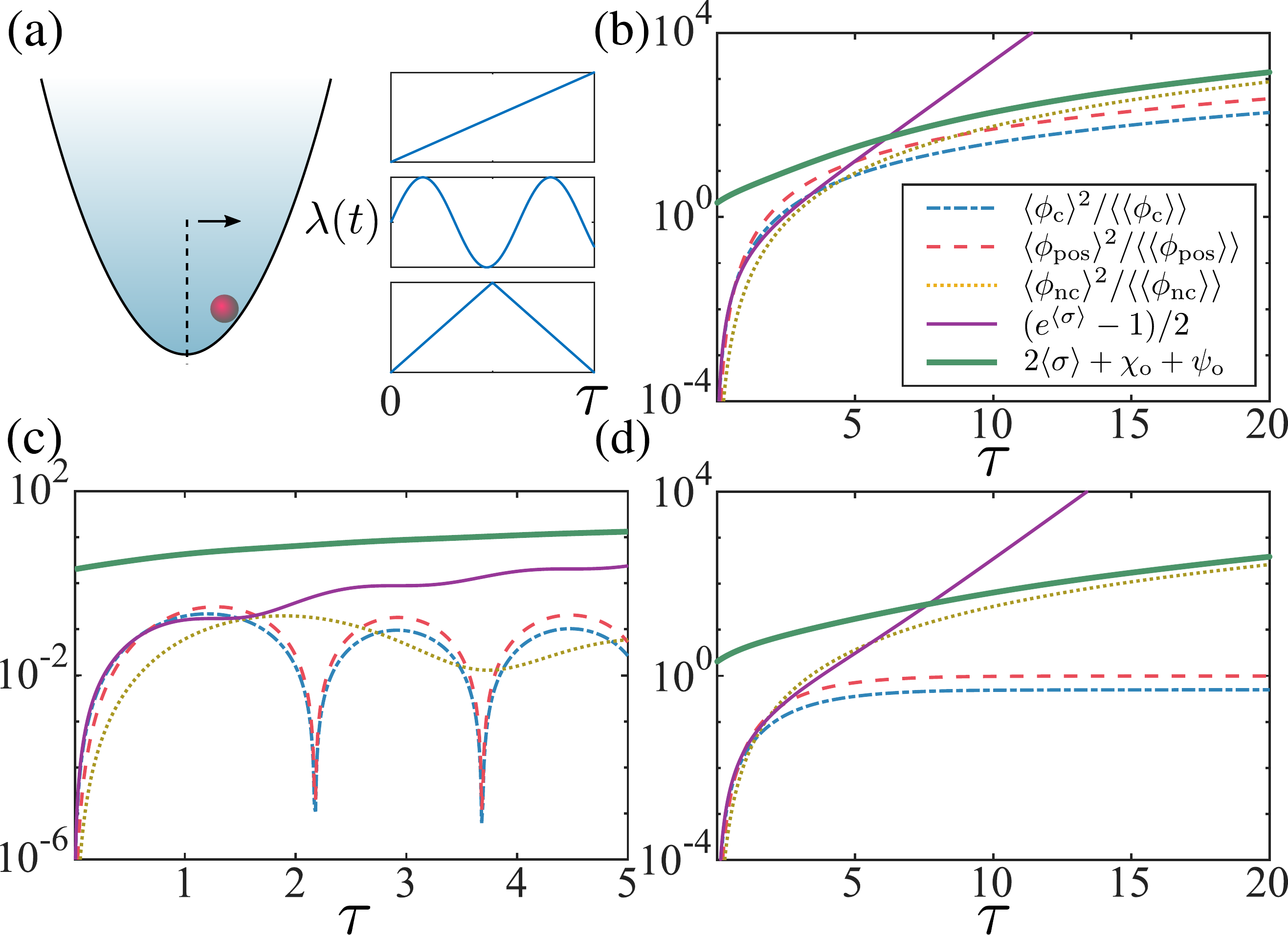}
	\protect\caption{(a) Schematic diagram of the dragged Brownian particle with the control protocol $\lambda$. Bound on fluctuations of the observables under (b) time-linear, (c) time-periodic, and (d) time-symmetric protocols [Eq.~\eqref{eq:bound.ex1}]. The dash-dotted, dashed, dotted, thin and thick solid lines represent the fluctuations $\avg{\phi_{\rm c}}^2/\Var{\phi_{\rm c}}$, $\avg{\phi_{\rm pos}}^2/\Var{\phi_{\rm pos}}$, $\avg{\phi_{\rm nc}}^2/\Var{\phi_{\rm nc}}$, the exponential bound $(e^{\avg{\sigma}}-1)/2$, and the derived bound $2\avg{\sigma}+\chi_{\rm o}+\psi_{\rm o}$, respectively. The derived bound is always satisfied, while the exponential bound is violated for all three cases. The observation time $\tau$ is varied, while the remaining parameters are fixed as $c=1,\alpha=1,\beta=2$, and $D=1$.}\label{fig:draggedBrownian}
\end{figure}

First, we investigate a dragged Brownian particle confined in a harmonic potential $U(x,\lambda)=c(x-\lambda)^2/2$, where $c>0$ is a constant [see Fig.~\ref{fig:draggedBrownian}(a) for illustration].
The total force is $F(x,\lambda)=-\pp_xU(x,\lambda)$, and the particle position is governed by the following equation:
\begin{equation}
\dot{x}=c(\lambda-x)+\xi.\label{eq:draggedBrownian}
\end{equation} 
We consider three cases: (a) a time-linear protocol $\lambda(t)=\alpha t$, (b) a time-periodic protocol $\lambda(t)=\alpha\sin(\beta t)$, and (c) a time-symmetric protocol
\begin{equation}
\lambda(t)=\begin{cases}
\alpha t, & 0\le t\le\tau/2,\\
\alpha(\tau-t), & \tau/2<t\le\tau,
\end{cases}
\end{equation}
where $\alpha$ and $\beta$ are positive constants.
We assume that the system is initially in equilibrium with the distribution $\rho_{\rm i}(x)=\exp(-cx^2/2D)$.
We consider three observables: a current representing the particle's displacement $\phi_{\rm c}(\Gamma)=x(\tau)-x(0)$, the final position $\phi_{\rm pos}(\Gamma)=x(\tau)$, and a noncurrent observable $\phi_{\rm nc}(\Gamma)=\int_0^\tau dt\,x$, which represents the area under the trajectory.
These observables satisfy the scaling condition with $\kappa=1$, i.e., $\phi(\theta\Gamma)=\theta\phi(\Gamma)$.
According to the derived bound [Eq.~\eqref{eq:odp.TUR}], inequality
\begin{equation}\label{eq:bound.ex1}
\frac{\avg{\phi}^2}{\Var{\phi}}\le 2\avg{\sigma}+\chi_{\rm o}+\psi_{\rm o}
\end{equation}
should be satisfied for all $\phi\in\set{\phi_{\rm c},\phi_{\rm pos},\phi_{\rm nc}}$.
All the terms in this bound can be analytically calculated in the following.

Let $\rho(x,t)$ be the probability distribution function of $x$ at time $t$.
Since the force is linear, the distribution is Gaussian, i.e., $\rho(x,t)=\mca{N}(x;\mu(t),\vartheta(t))$, where $\mu(t)$ and $\vartheta(t)$ are the mean and variance, respectively.
The initial conditions are $\mu(0)=0,~\vartheta(0)=D/c$.
From the Fokker--Planck equation, we obtain
\begin{equation}
\dot{\mu}(t)=c\bras{\lambda(t)-\mu(t)},\quad\vartheta(t)=\frac{D}{c}.\label{eq:diff.mu}
\end{equation}
Solving the differential equation [Eq.~\eqref{eq:diff.mu}] with respect to $\mu(t)$, we obtain
\begin{equation}
\mu(t)=c\int_0^tds\,e^{-c(t-s)}\lambda(s).
\end{equation}
Using the Laplace transform, the analytical solution of Eq.~\eqref{eq:draggedBrownian} can be expressed as
\begin{equation}
x(t)=\mu(t)+x_0e^{-ct}+\int_0^tds\,e^{-c(t-s)}\xi(s).\label{eq:odp.sol.x}
\end{equation}
From Eq.~\eqref{eq:odp.sol.x}, we obtain
\begin{equation}
\avg{[x(t)-\mu(t)][x(t')-\mu(t')]}=\frac{D}{c}e^{-c|t-t'|}.
\end{equation}
The observable averages can be analytically calculated, i.e., $\avg{\phi_{\rm c}}=\avg{\phi_{\rm pos}}=\mu(\tau)$ and $\avg{\phi_{\rm nc}}=\int_0^\tau dt\,\mu(t)$.
Analogously, the variances of the observables were obtained as $\Var{\phi_{\rm c}}=2D(1-e^{-c\tau})/c$, $\Var{\phi_{\rm pos}}=D/c$, and $\Var{\phi_{\rm nc}}=2D(e^{-c\tau}+c\tau-1)/c^3$.
The terms in the bound can be analytically calculated as
\begin{align}
\avg{\sigma}&=\frac{c^2}{D}\int_0^\tau dt\,(\lambda(t)-\mu(t))^2,\\
\chi_{\rm o}&=2c\tau-\frac{c^2}{2D}\int_{0}^{\tau}dt\bras{4\mu(t)^2+3\lambda(t)^2-8\lambda(t)\mu(t)},\\
\psi_{\rm o}&=2.
\end{align}
We illustrate the bound [Eq.~\eqref{eq:bound.ex1}] in Fig.~\ref{fig:draggedBrownian}(b)--(d), where the derived bound is always satisfied and the exponential bound is violated.
As seen, for short observation times, a gap exists between the bound and the fluctuation.
However, as $\tau$ is increased, the gap is reduced and the bound becomes tight.

\subsection{Brownian gyrator}
\begin{figure}[t]
	\centering
	\includegraphics[width=8.5cm]{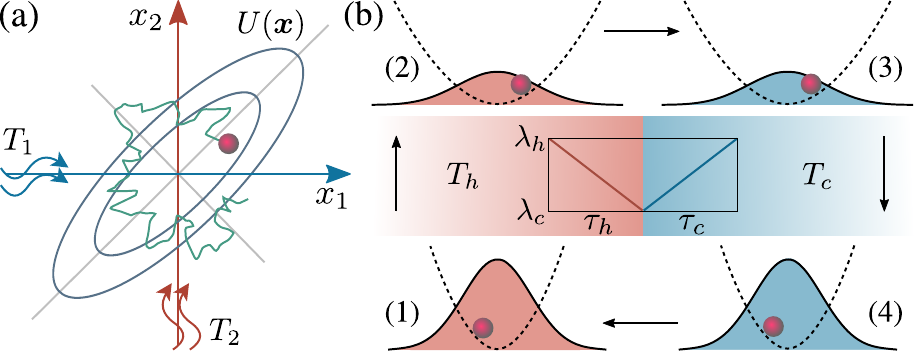}
	\protect\caption{Schematic diagrams of (a) the Brownian gyrator and (b) the stochastic underdamped heat engine. A cyclic period consists of four steps: isothermal expansion for a time $\tau_h$ [$(1)\to(2)$], instantaneously cooling the heat bath to temperature $T_c$ [$(2)\to(3)$], isothermal compression for a time $\tau_c$ [$(3)\to(4)$], and instantaneously heating the heat bath to temperature $T_h$ [$(4)\to(1)$]. The solid and dashed lines represent the probability distribution $\rho(x,t)$ and the potential $U(x,\lambda)$, respectively.}\label{fig:diagrams}
\end{figure}

\begin{figure}[t]
	\centering
	\includegraphics[width=8.5cm]{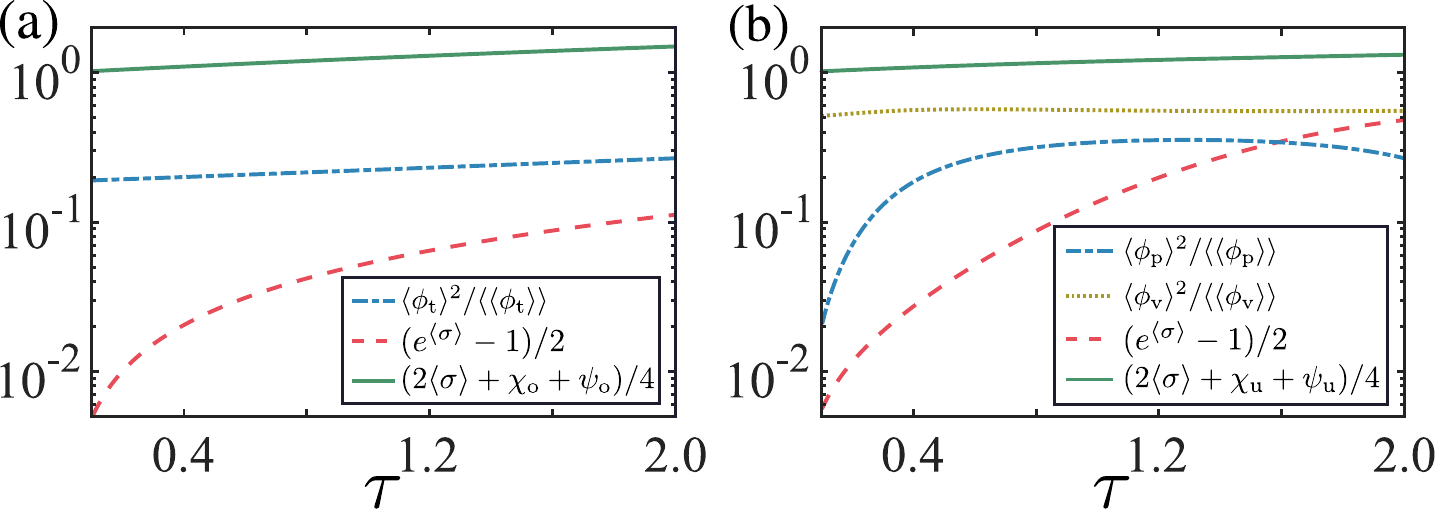}
	\caption{(a) Bound on the fluctuation of the accumulated torque [Eq.~\eqref{eq:bound.ex2}]. The system is in steady state, wherein the distribution is Gaussian. The dash-dotted, dashed, and solid lines represent the fluctuation $\avg{\phi_{\rm t}}^2/\Var{\phi_{\rm t}}$, the exponential bound $(e^{\avg{\sigma}}-1)/2$, and the derived bound $(2\avg{\sigma}+\chi_{\rm o}+\psi_{\rm o})/4$, respectively. The derived bound is always satisfied, while the exponential bound is not. The parameters are $\alpha=\pi/4$, $\gamma_1=\gamma_2=10$, $u_1=1$, $u_2=4$, $T_1=1$, and $T_2=4$. (b) Bound on fluctuations of the power output and the accumulated kinetic energy [Eq.~\eqref{eq:bound.ex3}]. The fluctuations $\avg{\phi_{\rm p}}^2/\Var{\phi_{\rm p}},\avg{\phi_{\rm v}}^2/\Var{\phi_{\rm v}}$, the exponential bound $(e^{\avg{\sigma}}-1)/2$, and the derived bound $(2\avg{\sigma}+\chi_{\rm u}+\psi_{\rm u})/4$ are depicted by the dash-dotted, dotted, dashed, and solid lines, respectively. The fluctuations exceed the exponential bound, while they are always smaller than the derived bound. The initial distribution $\rho_{\rm i}$ is a centered Gaussian with variances $\avg{x^2}=10,\avg{v^2}=1$, and $\avg{xv}=0$. The parameters are $m=1,\gamma_c=\gamma_h=0.1,T_c=1,T_h=4,\lambda_c=0.1,\lambda_{h}=2$, and $\tau_c=\tau_h=\tau/2$.}
	\label{fig:result}
\end{figure}

Next, we study a Brownian gyrator \cite{Filliger.2007.PRL}, which is a minimal microscopic heat engine that has recently been experimentally realized in an electronic and a colloidal system \cite{Chiang.2017.PRE,Argun.2017.PRE}.
The system consists of a particle with two degrees of freedom, $\bm{x}=[x_1,x_2]^\top$, trapped in an elliptical harmonic potential $U(\bm{x})=[u_1\bra{x_1\cos\alpha+x_2\sin\alpha}^2+u_2\bra{-x_1\sin\alpha+x_2\cos\alpha}^2]/2$,
where $u_1,u_2>0$ are stiffnesses along its principal axes, and $\alpha$ is the rotation angle.
The particle is simultaneously in contact with two heat baths at different temperatures acting in perpendicular directions [Fig.~\ref{fig:diagrams}(a)].
The particle position follows overdamped Langevin equations,
\begin{equation}
\gamma_i\dot{x}_i=-\pp_{x_i}U(\bm{x})+\xi_i,~(i=1,2),
\end{equation}
where $\gamma_i$ is the friction coefficient and $\xi_i$ is the zero-mean Gaussian white noise with covariance $\avg{\xi_i(t)\xi_j(t')}=2\delta_{ij}\gamma_iT_i\delta(t-t')$.
Here, $T_1\neq T_2$ are the temperatures of the heat baths.
In generic cases, i.e., $u_1\neq u_2$, the potential is asymmetric and a systematic gyrating motion of the particle around the potential minimum is induced due to the flow of heat.
The interesting observable is the accumulated torque exerted by the particle on the potential
\begin{equation}
\phi_{\rm t}(\Gamma)=\int_0^\tau dt\bras{x_1\pp_{x_2}U(\bm{x})-x_2\pp_{x_1}U(\bm{x})}.
\end{equation}
This observable is time-symmetric; thus, all TURs previously reported cannot be applied.
Since $\phi_{\rm t}(\theta\Gamma)=\theta^2\phi_{\rm t}(\Gamma)$, the following bound on torque fluctuation should be satisfied:
\begin{equation}
\frac{\avg{\phi_{\rm t}}^2}{\Var{\phi_{\rm t}}}\le\frac{\avg{\sigma}}{2}+\frac{\chi_{\rm o}+\psi_{\rm o}}{4}.\label{eq:bound.ex2}
\end{equation}
We illustrate Eq.~\eqref{eq:bound.ex2} in Fig.~\ref{fig:result}(a).
The fluctuation $\avg{\phi_{\rm t}}^2/\Var{\phi_{\rm t}}$ is numerically evaluated, while $\avg{\sigma}$, $\chi_{\rm o}$, and $\psi_{\rm o}$ are analytically calculated.
As seen, the bound is always satisfied when the observation time $\tau$ is varied.
Positive entropy production is needed to generate a nonzero torque.
However, the fluctuation cannot be bounded solely by entropy production, even with the exponential bound $(e^{\avg{\sigma}}-1)/2$.

\subsection{Stochastic underdamped heat engine}
Lastly, we consider a stochastic underdamped heat engine comprising a particle trapped in a harmonic potential $U(x,\lambda)=\lambda x^2/2$ \cite{Dechant.2017.EPL} [see Fig.~\ref{fig:diagrams}(b)].
The particle is embedded in a heat bath, whose temperature $T$ is cyclically varied to operate the system as a heat engine.
Its dynamics are described using the Langevin equation, 
\begin{equation}
m\dot{v}=-\gamma v-\lambda x+\xi,
\end{equation}
where the noise variance is $\avg{\xi(t)\xi(t')}=2\gamma T\delta(t-t')$.
We employ a time-linear protocol \cite{Blickle.2012.NP}
\begin{equation}
\lambda(t)=\begin{cases}
\lambda_{h}+(\lambda_{c}-\lambda_{h})t/\tau_h, & 0\le t<\tau_h,\\
\lambda_{c}+(\lambda_{h}-\lambda_{c})(t-\tau_h)/\tau_c, & \tau_h\le t<\tau,
\end{cases}
\end{equation}
where $\tau_h,\tau_c$ are the coupling times to the hot and cold heat baths, respectively, and $\tau=\tau_h+\tau_c$ is the total observation time.
The work $w$ exerted on the particle during the period is equal to $w(\Gamma)=\int_0^{\tau}dt\,\pp_\lambda U(x,\lambda)\dot{\lambda}$.
We consider two observables: the power output $\phi_{\rm p}=-w/\tau$ and the accumulated kinetic energy $\phi_{\rm v}=\int_0^{\tau}dt\,v^2$.
Since $\phi(\theta\Gamma)=\theta^2\phi(\Gamma)$, fluctuations of these observables are bounded as
\begin{equation}
\frac{\avg{\phi}^2}{\Var{\phi}}\le\frac{\avg{\sigma}}{2}+\frac{\chi_{\rm u}+\psi_{\rm u}}{4}\label{eq:bound.ex3}
\end{equation}
for $\phi\in\set{\phi_{\rm p},\phi_{\rm v}}$.
We assume that the initial distribution $\rho_{\rm i}(x,v)$ is Gaussian and illustrate Eq.~\eqref{eq:bound.ex3} in Fig.~\ref{fig:result}(b).
As shown, the derived bound is always satisfied, while the fluctuations cannot be constrained by the exponential bound.

We illustrate the implication of our results for the power output of heat engines.
The original TUR has been exploited to derive a bound on the fluctuation of power output in steady-state heat engines \cite{Pietzonka.2019.PRL}.
It indicates that a steady-state heat engine operating with Carnot's efficiency $\eta_C=1-T_c/T_h$ and delivering work with a finite fluctuation is impossible.
However, our bound does not imply this consequence in the same manner as the original bound.
It has been shown that one can construct a cyclic Brownian heat engine operating with efficiency asymptotically close to $\eta_C$ at nonzero power output with vanishing fluctuations \cite{Holubec.2018.PRL}.
Our bound is applicable to such an engine and such arbitrary heat engines described using Langevin dynamics.

\section{Conclusion}
Based on information theory, we have derived bounds for both current and noncurrent observables in overdamped and underdamped regimes.
These bounds universally hold for arbitrary protocols and arbitrary initial distributions.
The results demonstrate that the fluctuations of observables are constrained not only by entropy production but also by a kinetic term.
In all studied examples, we have shown that the exponential bound of entropy production, $(e^{\avg{\sigma}}-1)/2$, cannot constrain the fluctuation.
For the case of multipartite systems where the dynamics of the target system are unidirectionally affected by other systems, we have proved a tighter bound.
This bound reveals that the fluctuations of observables in the target system are not constrained by the dissipation costs of other systems but by the information flow between them and the target system.

Our results serve as a useful tool for estimation tasks in general Langevin systems.
Information inequalities have successfully been applied to derive many important thermodynamic bounds, such as the sensitivity-precision trade-off \cite{Hasegawa.2019.PRE}, a quantum TUR \cite{Guarnieri.2019.PRR}, and the speed limit \cite{Ito.2018.arxiv}.
Extending our approach to other classical and quantum systems or finding a hyper-accurate observable \cite{Busiello.2019.arxiv} would be interesting.

\section*{Acknowledgment}
This work was supported by Ministry of Education, Culture, Sports, Science and Technology (MEXT) KAKENHI Grant No.~JP16K00325 and No.~JP19K12153.

\appendix

\section{Derivation of the bounds for a full system}\label{app:deriv.fullsys}
To obtain Eqs.~\eqref{eq:odp.TUR} and \eqref{eq:udp.TUR}, we employ information-theoretic inequality with the perturbation technique \cite{Dechant.2018.JSM}.
We modify the force in the original system with a perturbation parameter $\theta$ and obtain new auxiliary dynamics.
For a given trajectory $\Gamma$, let $\mca{P}_\theta(\Gamma)$ denote the path probability of observing $\Gamma$ in the auxiliary dynamics.
According to the Cram{\'e}r--Rao inequality \cite{Hasegawa.2019.PRE}, the precision of the observable $\phi$ is bounded by the Fisher information as
\begin{equation}
\frac{(\pp_\theta\avg{\phi}_\theta)^2}{\Var{\phi}_\theta}\le\mca{I}(\theta).\label{eq:CRI.ine}
\end{equation}
Here, $\mca{I}(\theta):=\avg{\bra{\pp_\theta\ln\mca{P}_\theta(\Gamma)}^2}_\theta=-\avg{\pp_\theta^2\ln\mca{P}_\theta(\Gamma)}_\theta$ is the Fisher information.
Inequality \eqref{eq:CRI.ine} can be proven by applying the Cauchy--Swartz inequality to $(\pp_\theta\avg{\phi}_\theta)^2$ as follows:
\begin{equation}
\begin{aligned}
\bra{\pp_\theta\avg{\phi}_\theta}^2&=\bra{\pp_\theta\int \mca{D}\Gamma\,\mca{P}_\theta(\Gamma)\phi(\Gamma)}^2\\
&=\bra{\int \mca{D}\Gamma\,\mca{P}_\theta(\Gamma)(\phi(\Gamma)-\avg{\phi}_\theta)\pp_\theta\ln\mca{P}_\theta(\Gamma)}^2\\
&\le\Var{\phi}_\theta\,\mca{I}(\theta).
\end{aligned}
\end{equation}
For overdamped systems, we consider the auxiliary dynamics, $\dot{x}=H_\theta(x,t)+\xi$, where
\begin{equation}
H_\theta(x,t)=\theta F(x/\theta,\lambda)+D(1-\theta^2)\frac{\pp_x\rho(x/\theta,t)}{\rho(x/\theta,t)}.
\end{equation}
Analogously, for underdamped systems, the dynamics are modified as $m\dot{v}=H_\theta(x,v,t)+\xi$, where
\begin{equation}
\begin{aligned}
H_\theta(x,v,t)=-\gamma v+\theta F(x/\theta,\lambda)+\frac{D}{m}(1-\theta^2)\frac{\pp_v\rho(x/\theta,v/\theta,t)}{\rho(x/\theta,v/\theta,t)}.
\end{aligned}
\end{equation}
When $\theta=1$, these auxiliary dynamics become the original ones.
The distributions of auxiliary dynamics in the overdamped and underdamped cases are $\rho_\theta(x,t)=\rho(x/\theta,t)/\theta$ and $\rho_\theta(x,v,t)=\rho(x/\theta,v/\theta,t)/\theta^2$, respectively.
In both cases, the observable average is scaled as $\avg{\phi}_\theta=\theta^\kappa\avg{\phi}$; thus, $\pp_\theta\avg{\phi}_\theta|_{\theta=1}=\kappa\avg{\phi}$.
The path probability using the pre-point discretization can be expressed via the path-integral representation as
\begin{equation}
\mca{P}_\theta(\Gamma)=\mca{N}_{\rm o}\rho_\theta(x(0),0)\exp\bra{-\int_0^\tau dt\frac{(\dot{x}-H_\theta(x,t))^2}{4D}}
\end{equation}
for the overdamped case and
\begin{equation}
\mca{P}_\theta(\Gamma)=\mca{N}_{\rm u}\rho_\theta(x(0),v(0),0)\exp\bra{-\int_0^\tau dt\frac{(m\dot{v}-H_\theta(x,v,t))^2}{4D}}
\end{equation}
for the underdamped case.
Here, $\mca{N}_{\rm o}$ and $\mca{N}_{\rm u}$ are terms independent of $\theta$.
Note that entropy production $\avg{\sigma}$ is $\int_0^\tau dt\int dx\,j(x,t)^2/D\rho(x,t)$ in overdamped systems and $\int_0^\tau dt\int dxdv\,j^{\rm ir}(x,v,t)^2/D\rho(x,v,t)$ in underdamped systems.
Here, $j^{\rm ir}(x,v,t)=-1/m[\gamma v+D/m\pp_v]\rho(x,v,t)$ is the irreversible probability current.
Consequently, performing simple algebraic calculations, one can show that $\mca{I}(1)$ is equal to $2\avg{\sigma}+\chi_{\rm o}+\psi_{\rm o}$ for the overdamped case and to $2\avg{\sigma}+\chi_{\rm u}+\psi_{\rm u}$ for the underdamped case.
By letting $\theta=1$ in Eq.~\eqref{eq:CRI.ine}, we obtain the uncertainty relations given in Eqs.~\eqref{eq:odp.TUR} and \eqref{eq:udp.TUR}.

\section{Bounds for multidimensional systems}\label{app:multidimen}
We consider $n$-dimensional systems with variables $\bm{x}=[x_1,x_2,\dots,x_n]^\top$.
We consider observables that satisfy the scaling condition: $\phi(\theta\Gamma)=\theta^\kappa\phi(\Gamma)$, where $\kappa>0$ is a real constant.
Specifically, we focus on three types of observables: a current observable $\phi(\Gamma)=\int_0^\tau dt\,\bm{\Lambda}_{\rm c}(\bm{x})^\top\circ\dot{\bm{x}}$, a noncurrent observable $\phi(\Gamma)=\int_0^\tau dt\,\Lambda_{\rm nc}(\bm{x})$, and a discrete-time observable $\phi(\Gamma)=\sum_{i}c_i\Lambda_{\rm nc}(\bm{x}(t_i))$, where $\bm{\Lambda}_{\rm c}(\bm{x})$ and $\Lambda_{\rm nc}(\bm{x})$ satisfy that $\bm{\Lambda}_{\rm c}(\theta\bm{x})=\theta^{\kappa-1}\bm{\Lambda}_{\rm c}(\bm{x})$ and $\Lambda_{\rm nc}(\theta\bm{x})=\theta^{\kappa}\Lambda_{\rm nc}(\bm{x})$.
By employing the same modified dynamics as in Appendix \ref{app:deriv.fullsys}, one can show that the probability currents and distribution function in the auxiliary dynamics are scaled as $\rho_\theta(\bm{x},t)=\rho(\bm{x}/\theta,t)/\theta^n,~j_{\theta}(\bm{x},t)=j(\bm{x}/\theta,t)/\theta^{n-1}$ for overdamped cases and $\rho_\theta(\bm{x},\bm{v},t)=\rho(\bm{x}/\theta,\bm{v}/\theta,t)/\theta^{2n},~j_\theta(\bm{x},\bm{v},t)=j(\bm{x}/\theta,\bm{v}/\theta,t)/\theta^{2n-1}$ for underdamped cases.
Consequently, it is easy to verify that $\avg{\phi}_\theta=\theta^\kappa\avg{\phi}$.
For $n$-dimensional overdamped systems described as
\begin{equation}
\dot{x}_i=F_i(\bm{x},\lambda)+\xi_i,~(i=1,\dots,n),
\end{equation}
the uncertainty relation reads
\begin{equation}
\frac{\avg{\phi}^2}{\Var{\phi}}\le\frac{1}{\kappa^2}\bra{2\avg{\sigma}+\chi_{\rm o}+\psi_{\rm o}},\label{eq:mul.odp.TUR}
\end{equation}
where the terms in the right-hand side of Eq.~\eqref{eq:mul.odp.TUR} are defined by
\begin{align}
\chi_{\rm o}&:=\int_0^\tau dt\int d\bm{x}\,\Lambda_{\rm o}(\bm{x},t)\rho(\bm{x},t),\\
\psi_{\rm o}&:=\Big\langle\Big(\sum_{i=1}^nx_i\pp_{x_i}\rho_{\rm i}/\rho_{\rm i}\Big)^2\Big\rangle_{\rho_{\rm i}}-n^2.
\end{align}
Here, $\Lambda_{\rm o}=\sum_{i=1}^n\bra{G_i^2-4F_iG_i-4D_i\pp_{x_i}G_i}/2D_i$ and $G_i=F_i+\sum_{j=1}^nx_j\pp_{x_j}F_i$.
Analogously, for $n$-dimensional underdamped systems described as
\begin{equation}
\dot{x}_i=v_i,~m\dot{v}_i=-\gamma_iv_i+F_i(\bm{x},\lambda)+\xi_i,~(i=1,\dots,n),
\end{equation}
the bound has the following form:
\begin{equation}
\frac{\avg{\phi}^2}{\Var{\phi}}\le\frac{1}{\kappa^2}\bra{2\avg{\sigma}+\chi_{\rm u}+\psi_{\rm u}},\label{eq:mul.udp.TUR}
\end{equation}
where the terms in the right-hand side of Eq.~\eqref{eq:mul.udp.TUR} are defined by
\begin{align}
\chi_{\rm u}&:=\int_0^\tau dt\int d\bm{x}d\bm{v}\,\Lambda_{\rm u}(\bm{x},\bm{v},t)\rho(\bm{x},\bm{v},t),\\
\psi_{\rm u}&:=\Big\langle\Big[\sum_{i=1}^n\bra{x_i\pp_{x_i}\rho_{\rm i}+v_i\pp_{v_i}\rho_{\rm i}}/\rho_{\rm i}\Big]^2\Big\rangle_{\rho_{\rm i}}-4n^2.
\end{align}
Here, $\Lambda_{\rm u}=\sum_{i=1}^n[(F_i-\sum_{j=1}^nx_j\pp_{x_j}F_i)^2-4\gamma_i^2v_i^2+8\gamma_iD_i/m_i]/2D_i$.

\section{Derivation of the bound for a subsystem}\label{app:deriv.subsys}
We consider the following auxiliary dynamics:
\begin{align}
\dot{x}&=\theta F_x(x/\theta,y,\lambda)+D_x(1-\theta^2)\frac{\pp_x\rho(x/\theta,y,t)}{\rho(x/\theta,y,t)}+\xi_x,\\
\dot{y}&=F_y(y,\lambda)+\xi_y,
\end{align}
Unlike the previous modifications, we change only the dynamics of the target system, $\mX$, and keep those of other systems, $\mY$, unchanged.
It can be verified that the distribution of this dynamics is scaled as $\rho_\theta(x,y,t)=\rho(x/\theta,y,t)/\theta$, while the probability currents are scaled as $j_{x,\theta}(x,y,t)=j_x(x/\theta,y,t)$ and $j_{y,\theta}(x,y,t)=j_y(x/\theta,y,t)/\theta$.
Subsequently, by applying the same procedure as in Appendix \ref{app:deriv.fullsys}, one can obtain Eq.~\eqref{eq:inter.TUR}.

%\bibliography{refs}

%

\end{document}